# A New Secure Network Architecture to Increase Security among Virtual Machines in Cloud Computing

Zakaria Elmrabet

National Institute of Posts and Telecommunications, Rabat Morocco
zakaria.elmrabet2@gmail.com

Hamid Elghazi

International University of Rabat, Rabat Morocco
hamid.elghazi@uir.ac.ma

Tayeb Sadiki

International University of Rabat, Rabat Morocco
tayeb.sadiki@uir.ac.ma

Hassan Elghazi

National Institute of Posts and Telecommunications, Rabat Morocco
elghazi@inpt.ac.ma

**Abstract.** Cloud computing is a new model of computing which provides scalability, flexibility and on-demand service. Virtualization is one of the main components of the cloud, but unfortunately this technology suffers from many security vulnerabilities. The main purpose of this paper is to present a new secure architecture of Virtual Network machines in order to increase security among virtual machines in a virtualized environment (Xen as a case study). First, we expose the different network modes based on Xen Hypervisor, and then we analyses vulnerabilities and security issues within these kind of environment. Finally, we present in details new secure architecture and demonstrate how it can face the main security network attacks.

## 1      Introduction

The Cloud Computing is a new model and computing paradigm which offer scalable on-demand services to consumer with greater flexibility and lesser infrastructure investment. Economic benefits of the cloud computing are the main factors that encourage consumers to adopt it, according to [1] 91% of organizations in US and Europe agreed that reduction in cost is a major reason for them to migrate to this new environment.

Virtualization is one of the main components of the cloud computing, this component is important because it enables multi-tenancy and on-demand use of scalable shared resources [2]. Virtualization means create a virtual version of a device or resources such as Server, Storage or Network [3] and implements it using a "Hypervisor ".

However, vulnerabilities in a virtual environment expose the security of the information stored in the cloud computing to many big challenges [4]. This vulnerability is classified on three main issues: availability, integrity and confidentiality of the information (or the CIA triad).

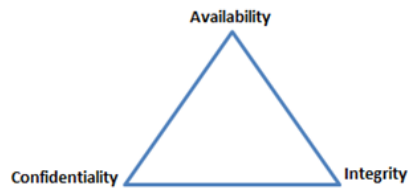

**Fig. 1.** The CIA Triad of information security

The CIA triad, are the three fundamental pillars of information security [5]. They are defined as:

- **Availability**: The data is available when needed.
- **Integrity**: The data is not modified without being detected.
- **Confidentiality**: The data remains undisclosed to unauthorized parties

Thus, any global solution for security problems within Cloud Computing shall take into account these three parameters. However, not every attack compromises every attribute of the triad, there are some attacks affect the three attributes, whereas some others only one or two attributes.

The table 1 shows the impact of the most known dangerous attacks in Cloud Computing on the availability, integrity and confidentiality of the stored information: spoofing attack, sniffing attack and the Mac flooding attack.

**Table 1.** Security Attributs affected by network attacks

|  | Spoofing/ Poisoning | Sniffing | Mac Flooding |
|---|---|---|---|
| Availability | X |  | X |
| Integrity | X |  |  |
| Confidentiality | X | X | X |

We believe that in the context of Cloud Computing and in order to guarantee security among virtual machines, hypervisor should ensure isolation of each virtual machine (VM) by using dedicated physical channel. However, and as presented previously hypervisor (In our case Xen Hypervisor) uses software channel to create bridge and in order to link virtual machines. In fact, software channels isolation is easy to be broken [6].

The main purpose of this paper is to propose a new architecture to increase security in virtual environment and especially in virtual network; the hypervisor used in our study is the Xen Hypervisor. The proposed architecture is composed of two main layers: Switch Layer and Firewall Layer. The purpose of the switch layer is to prevent from sniffing attack, whereas the firewall layer will prevent from spoofing and Mac flooding attack.

In this paper, we start first by introducing the virtual network modes in Cloud Computing. Then, in the next sections we expose a critical analysis of security issues within these modes. Finally, we present in more details our new architecture model.

## 2 Related Work

In order to secure the communication among virtual machines the Xen Hypervisor environment, [12] proposes a model composed of three layers: Routing layer, Firewall and shared Network.

The routing layer connects the physical network and creates a logical dedicated channel in order to establish the communication between the virtual network and the physical network. The administrator assigns a set of unique static IDs to each shared network, these ID are stored in a configuration file and they could be used to monitor the source of packets sent from each shared network [12].

The second layer is the Firewall layer; the purpose of this layer is to prevent spoofing attacks between virtual machines belongs to a different shared network, by identifying the network ID specified in the configuration file [12].

The third and the last layer in the proposed model is the shared network layer, in this layer the authors assume that the VMs belong to a same virtual shared network are trustful to each other and they are working for the same company or organization [12].

The [12] model is a good approach to increase security among virtual machines within a virtual network. However, it can prevent only spoofing attack. Other attacks, for instance: Sniffing, and Mac flooding still exist and a malicious person could use a virtual machine to launch these kinds of attacks. In fact, it could affect the availability and the confidentiality of the information hosted in the cloud.

## 3    Critical Analysis of Vulnerabilities in Virtual Network

### 3.1    Virtual Network

Virtual network is a technique used to create independent or isolated logical network within a shared physical network. Existing hypervisors (such as Xen and Vmware) offer this mechanism to share the access of physical network [12]. Xen is open source software and we will use it as an example of application of our approach.

Xen is a project developed in 2003 at the University of Cambridge Laboratory. There are both commercial and free versions of Xen.it is a Hypervisor that provides a platform for running multiple instance of the operating system on one physical hardware. It supports several operating systems such as: Linux, FreeBSD and also Microsoft Windows.in this paper the free version of Xen will be used [11].

Xen Hypervisor refers to each virtual machine as a domain, and there are two types of domains: the first one called Domain0 (dom0) is a privileged domain that can access the hardware resources and also contains some tools for managing other domains in other word other virtual machines, while the second type is called domU, and it is an unprivileged virtual machines that created and managed by dom0 [11].

All requests from domU instances for hardware are passed via dom0, and then they will be forwarded to the actual hardware. The figure below illustrates an example of 3 virtual machines: dom0, dom1 and dom2 running on Xen hypervisor.

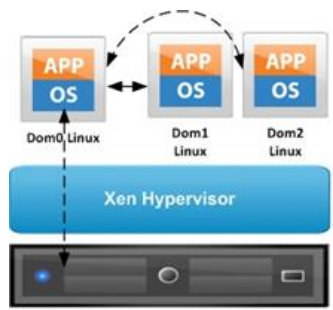

**Fig. 2.** Three virtual machines: dom0, dom1 and dom2 running on Xen Hypervisor

Xen Hypervisor provides different networking modes to configure virtual network:

a.    **Bridged networking**

Bridge is a technique used to connect two LANs (Local area network) together and forwards frames using their MAC (media access control) address. It is the default option for Xen networking.

Before forwarding frames, the bridge checks its bridge table where MAC addresses are stored. In order to fill up this table with the MAC addresses the bridge uses broadcasting.

The following figure shows an example of two virtual machines (dom0, domU:1 and domU:2 ) connecting to the bridge.

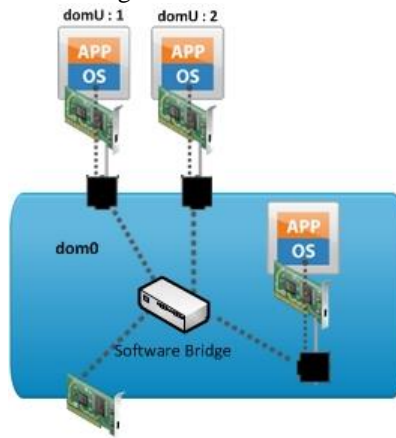

**Fig. 3.** Xen Bridged Networking

### b. Routed Networking

Routed networking is the second mode offered by Xen Hypervisor, is a technique that uses IP addresses to send and receive network traffic from one segment to another. As shown in the figure below, dom0 in Xen acts as a default gateway for the other domains to communicate with the outside Word.

In addition, the domU machines are visible from the outside and can be directly accessed via the default gateway in this case dom0.

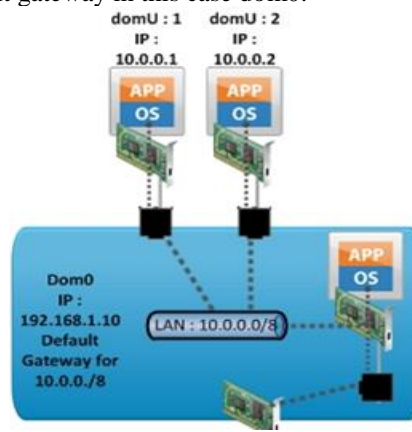

**Fig. 4.** Xen Routed Networking

c. **Virtual Local Area Network (VLAN) with Network Address Translation (NAT).**

The VLAN is the third mode for configuring virtual network in Xen Hypervisor. In this mode the domU machines are created on a Private LAN, and they use the dom0 as a default gateway to reach the outside.

The mean difference between this mode and Routed mode, in the first one the domU machines are visible from the outside while in the second mode the domU machines are hidden and protected from the outside.

### 3.2 Vulnerabilities in Virtual Network

In order to guarantee security among virtual machines within virtual network, hypervisor should ensure isolation. The good secure way to isolate each virtual machine (VM) is by using dedicated physical channel as presented previously hypervisor uses software to create bridge, route or NAT. Thus, isolation is easy to be broken [12].

Bellows are presented the security challenges related to virtual network and especially in Xen Hypervisor:

- **Spoofing and Poisoning.**

Arp (Address Resolution protocol) is a protocol used in the local Network to resolve an IP Address into MAC (Medium Access Controllers) address. A Virtual Machine (VM) invokes an Arp resolution when it needs a MAC address of a new IP, and then the result is saved in its cache [7].

Arp poisoning is a malicious technique used to modify the association between an IP address and its corresponding MAC address [7]. Arp poisoning utilizes Arp Spoofing, a VM sends a spoofed ARP message with a forged IP address to the "Virtual Switch», once the Virtual Switch receives the request it will dynamically update its cache. As a result, frames intended for the legitimate VM can be mistakenly sent to the Attacker VM.

To illustrate the processes of spoofing within a virtual network, assuming that a VM-A would communicate with VM-B and the attacker VM-C will launch a spoofing attack in order to capture the communication. The figure 5 (a) illustrates the state of the routing table before launching the attack.

Once the routing table records the IP address, MAC address and the port of each virtual machine in the virtual network, the attacker virtual machine VM-C sends an ARP request with a forged IP addresses, for the first time with the same IP address as VM-A and for a second time with the same IP address as VM-B, and then the virtual route will update the routing table with the new information coming from the attacker virtual machine VM-C as illustrated in figure 5 (b). In fact, any traffic coming to or from the VM-A or VM-B would be mistakenly sent to VM-C instead.

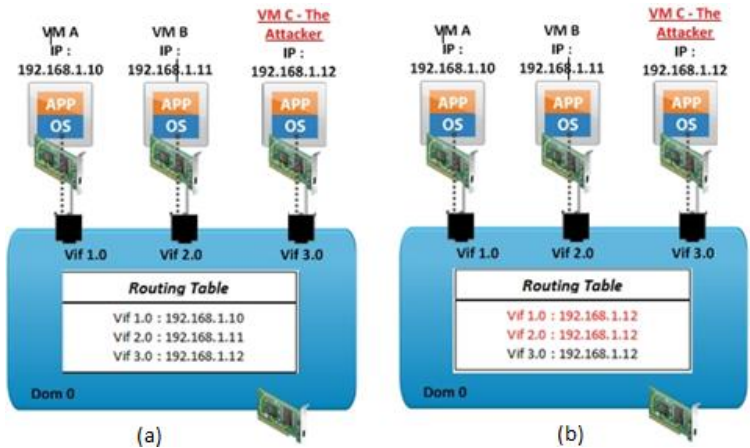

**Fig. 5.** Spoofing Attack in Xen Hypervisor (a) Content of the routing table before launching the attack, (b) Content of the routing table before launching the attack

The spoofing attack can be used to conduct other attacks resultant a serious damage, such as:

- **Denial-of-service (Dos)** : it is an attack where the malicious person launches a spoofing attack, and then he/she chooses to    block the frames passing through he/she machine instead of just sniffing or modifying its contents.as a result the regulars  virtual machines  could not communicate with each other.
  - **Man-in-the –middle** : it is an attack where the malicious person launches a spoofing attack, and then he/she chooses to listen or to modify the communication between two virtual machines, and unfortunately  these ones believe that they communicate directly with each other.
  - **Session Hijacking**: it is an attack where the malicious person stole a cookie after launching a spoofing attack. Recall that cookies are used by the client to maintain a session on web servers. Therefore, if the attacker stole the cookie, he/she could connect easily to the web server with the legitimate user's privileges, and he/she could access to sensitive data.

In Xen Hypervisor, the route mode plays role as a "Virtual switch"[12], therefore a malicious VM can launch an ARP spoofing  attack against the "virtual switch" in order to redirect the traffic to its machine and then conduct other malicious attacks.

- **Sniffing.**

Sniffing or listen to the traffic, it is an attack where the attacker insert itself between two communicating hosts in order to capture the frames and then retransmit it [8][9][10].

In Xen Hypervisor, Bridge mode is a technique used to connect two LAN (Local Area Network) together and it forwards frames using their MAC (media access control) address [11]. Therefore the bridge plays a role as a "Virtual Hub" for each seg-ment [12], in which a malicious VM is able to sniff easily the virtual network by first configuring the "promiscuous mode" of its network card and then using a free sniff tool such as "tcpDump" and "Wireshark"[13].

Recall that the "promiscuous mode" refers to an operation mode where a network card is configured to accept every packet transmitted in the network whether they are addressed to this network card or not.

To illustrate the processes of sniffing within a virtual network in this case: Xen Hypervisor, assuming that VM-A would communicate with VM-B and the Attacker VM-C will sniff the communication.so the VM-A sent a message to a VM-B, the message will be received first by the bridge before forwarding it to its destination VM-B, so once the bridge receives the message, it retransmit it to the VM-B and as VM-C is placed on the same segment as VM-B it is able to capture or to sniff the message (as illustrated in figure 6). As result VM-C could have access to a sensitive and confidential data.

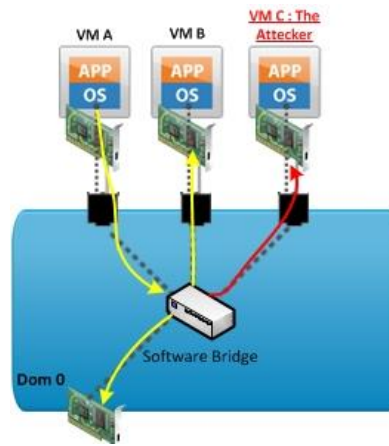

**Fig. 6.** Sniffing attack in Xen Hypervisor

- **Mac Flooding**.

The Content-Addressable Memory (CAM) table stores MAC addresses in the switch, and it is fixed size. This attack consists of flooding a switch with MAC addresses using forged ARP packets until the CAM table is full. Then, the switch operates as a hub and starts broadcasting the traffic without a CAM entry [14].

In Xen Hypervisor, the route mode plays role as a "Virtual switch"[12], therefore a malicious VM can launch a MAC flooding attack against the "virtual switch" in order to fill up it CAM table and then sniff the traffic coming from and going to other VMs.

### 3.3 Security vulnerabilities on Virtual Network Mode

The three modes available on Xen Hypervisor to configure network are: Bridge Mode, Route Mode and NAT, and as described above these modes are vulnerable. The table below resumes the security vulnerabilities related to each virtual network mode of Xen Hypervisor. The sign (-) means that we can launch an attack in that mode, and the sign (+) means we cannot lunch an attack in that mode.

**Table 2.** Vulnerabilities in virtual network

| Network Mode\Attack | Routed Mode | NAT Mode | Bridged Mode |
|---|---|---|---|
| Spoofing | - | - | - |
| Sniffing | + | + | - |
| Mac Flooding | - | - | - |

From the table 2 we think that a new architecture shall be defined to protect virtual network from attacks. The main purpose of this paper is to provide architecture inspired from classical network security. This kind of architecture proved their efficiency and confidence in classical network, thus they can be reproduced in the context of virtual network.

## 4   A New Architecture Model

In order to increase security among virtual machines, we propose a new architecture model to prevent users from network attacks. Especially, from spoofing, sniffing and mac flooding attack. This architecture is composed of three layers: VLans Layer, Virtual Switch layer and firewall layer as shown in figure 7.

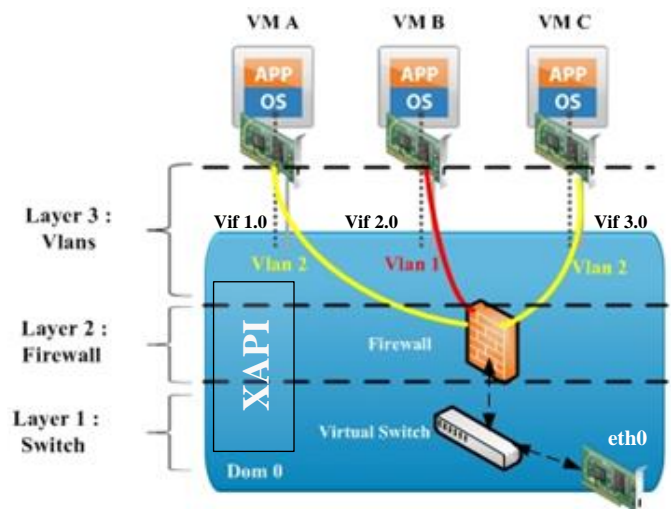

**Fig. 7.** A secure virtual network model

- **Layer 1: Virtual Switch,** the main purpose of this layer is to prevent sniffing attack. First, it allows virtual machines to communicate with the outside network and also permits virtual machines belong to the same VLAN to communicate securely with each other. The virtual machines belong to the same organization are assigned by the administrator to a specific VLAN, in fact the communication among virtual machines within this VLAN could not be sniffed form other VLANs.

Open virtual Switch [16] is an open source tool and it could be used with Xen Hypervisor to implement a virtual switch.

The integration of Open vSwitch with the Hypervisor is similar to a bridged configuration; however, instead of connecting each virtual network card (vif) to a bridge the open virtual switch will be used. As an example to set up a virtual switch (in this case the Open Virtual switch) we could use the following steps :

- After the installation, we create a new bridge named vswith0:

```
ovs-vsctl add-br vswith0
```

- And we attach the physical network interface card (eth0) to the new bridge (vswith0) :

```
ovs-vsctl add-port vswith0 eth0
```

- **Layer 2: Virtual Firewall,** will be used in order to prevent spoofing, poisoning and MAC flooding attack a firewall is crucial. At this level, we define a set of security policies mainly including: any packet send from one VLAN and attempts to reach another VLAN will be dropped. Once the virtual switch records information about all virtual machines on the virtual network in its CAM table, any packet tries to modify or to fill up this table will be dropped.

   Iptable [15] an open source tool could be used with Xen Hypervisor to implements these policies.

- **Layer 3 : Vlans**, provides a logical segmentation of switch ports, allowing communication as if all ports were on the same physical LAN segment. Limiting the broadcast traffic to a subset of the switch ports saves significant amounts of network bandwidth and also to increase security among users by preventing some attacks such as: sniffing[17][18].

   As an open source tool, Open virtual switch could be used to either implement a virtual switch and to configure Vlans [19].

   As an example to configure VLANs on open Virtual Switch, we could perform the following configuration [19]:

   - Create an Open Virtual Switch

   ```
   ovs-vsctl add-br br0
   ```

   - Add the physical network interface (eth0) to the virtual switch (br0)

   ```
   ovs-vsctl add-port br0 eth0
   ```

   - Add VM A (which has the virtual interface vif1.0) and VM C (which has the virtual interface vif3.0) on VLAN 2. This means that traffic coming into virtual switch from VM A or VM C will take the tag number 2 and will be considered part of VLAN 2:

```
ovs-vsctl add-port br0 vif1.0 tag=2
ovs-vsctl add-port br0 vif3.0 tag=2
```

- Add VM B (which has the virtual interface vif 2.0 ) on VLAN 1.this means that traffic coming to the virtual switch from VM B will belong to the VLAN 1.

```
ovs-vsctl add-port br0 vif2.0 tag=1
```

As a result, the virtual machines belong to the same Vlan, for instance VM A and VM C witch belong to the VLAN 2 should succeed, however virtual machines belong to a different VLANs, for example VM A and VM B should not succeed. Therefore the isolation of traffic is assured by this method.

In order to manage all aspect of Xen, including VMs, Storage, and especially networking, we have used XAPI (Xen Server API), it provides an external interface for configuring the system and also enables the hypervisor to establish a communication with other plug-in such as Open Virtual Switch[20].

In order to evaluate the efficiency of this new model regarding attack within a virtual network and specially: sniffing, spoofing, poisoning and MAC flooding attacks. The table below presents a comparison between the proposed model in this paper and the existing modes.

**Table 3.** Comparaison between the proposed mode and the existing virtual network modes in Xen Hypervisor in term of security

| Network Mode / Attack | Routed Mode | NAT Mode | Bridged Mode | Proposed Mode |
|---|---|---|---|---|
| Spoofing | - | - | - | + |
| Sniffing | + | + | - | + |
| Mac Flooding | - | - | - | + |

## 5  Conclusion

The main purpose of this paper is to improve security in the cloud and especially in the virtual network, which constitute the main component of the cloud. Through our work we presented a critical security analysis of the different mode of virtual network (based on Xen Hypervisor). Then we expose the main vulnerability related to those modes. Finally, we described in more details a new architecture witch composed of three layers, Vlans, Firewall and Switch. This new architecture can prevent from some attacks such as: spoofing, sniffing and mac flooding.

As future work, we will implement the new architecture in Xen hyervisor, with the Firewall Iptable and the Open virtual Switch, in order to evaluate its efficiency in term of security and performance.

# References

1. Chirag M.; Patel D.; et al., A survey on security issues and solutions at different layers of Cloud computing, The Journal of Supercomputing, Volume 63, Issue 2, pp 561-592. Springer (2013).
2. Salah, K.; et al., "Using Cloud Computing to Implement a Security Overlay Network," Security & Privacy, vol.11, no.1, pp.44,53, Jan.-Feb. IEEE (2013)
3. Vangie B. http://www.webopedia.com/TERM/V/virtualization.html (website visited April 2015)
4. National vulnerability database version 2.2.NIST. http://web.nvd.nist.gov/view/vuln/searchresults?query=virtual&search%_type=all&cves=on (website visited mars 2015)
5. Mariam K., A Methodology for Cloud Security Risks Management, Cloud Computing Challenges, Limitations and R&D Solutions, pp. 75-104. Springer (2014).
6. Grover, J.; Shikha; Sharma, M., "Cloud computing and its security issues — A review," Computing, Communication and Networking Technologies (ICCCNT), 2014 International Conference on ,pp.1,5. , IEEE (2014).
7. Bruschi, D.; Ornaghi, A.; Rosti, E., "S-ARP: a secure address resolution protocol," Computer Security Applications Conference, 2003. Proceedings. 19th Annual, pp.66,74, 8-12.,IEEE (2003).
8. Chin, Tan Saw, Singh Y P. Single-hop wavelength assignment using an ant algorithm in WDM MESH network.WSEAS Transactions on Computers.Vol.5, No.7, 2006, pp. 294-300.
9. Wenbing Zheng, Chenzhong LI, "AN Algorithm Against Attacks Based on ARP Spoofing", Journal of Southern Yangtze University(Natural Science Edition), Vol.2, No.6, 2003, pp. 167-1696.
10. Z. H. Tian, B. X. Fang, B. Li, et al. Avulnerability-driven approach to active alert verification for accurate and efficient intrusion detection. WSEAS Transactions on Communications.Vol.4, No.10, 2005, pp. 1002-1009.
11. P. Chaganti, Xen Virtualization, 2007, pp. 74.
12. Hanqian Wu; et al., "Network security for virtual machine in cloud computing," Computer Sciences and Convergence Information Technology (ICCIT), 2010 5th International Conference on , vol., no., pp.18,21, IEEE (2010).
13. CERT Training and Education, Carnegie Mellon University, 2009 http://science.hamptonu.edu/compsci/docs/iac/packet_sniffing.pdf
14. Hayriye A.; Sven K.,Henry L.; et al., "Securing Layer 2 in Local Area Networks", Networking - ICN 2005, 4th International Conference on Networking, Reunion Island, France, Proceedings, Part II,pp 699-706. Springer (2005).
15. http://www.net security.org/software.php?id=4 (website visited March 2015)
16. Open vswitch http://openvswitch.org/ (website visited April 2015)
17. http://www.omnisecu.com/cisco-certified-network-associate-ccna/advantages-of-vlan.php (website visited April 2015)
18. Vmweare Virtual Networking Concept (2007). http://www.vmware.com/files/pdf/virtual_networking_concepts.pdf
19. Vlan configuration. http://openvswitch.org/support/config-cookbooks/vlan-configuration-cookbook/(website visited April 2015)
20. Justin p.; et al., "Virtual Switching in an Era of Advanced Edges", http://benpfaff.org/papers/adv-edge.pdf (website visited July 2015)